\documentclass[aps,prb,reprint]{revtex4-2}
\usepackage{graphicx}
\usepackage{graphics}
\usepackage{amsfonts}
\usepackage{amsmath}
\usepackage{amssymb}

\begin{document}

\title{Alternating current Hanle effect as poor man's paramagnetic resonance}

\author{Ya.\ B. Bazaliy}
  \email{bazaliy@mailbox.sc.edu}
  \affiliation{University of South Carolina, Columbia, SC 29208, USA}

\date{\today}

\begin{abstract}
It is shown that in spin injection experiments the interplay between external magnetic field and alternating current can be observed already on a single ferromagnet/normal metal interface. The interface resistance is predicted to exhibit prominent features whenever the frequency of spin precession in the applied field becomes equal to the frequency of the driving current. Using these features, material-specific g-factors of electrons in normal metals can be measured with less effort, albeit also with less precision.
\end{abstract}

\maketitle

\section{Introduction}

In heterostructures made of ferromagnetic (F) and normal metal (N) parts, electric currents flowing through the F/N interfaces inject spins into normal regions and cause non-equilibrium spin accumulation. The latter is known to induce extra voltage on each F/N interface or, in other words, leads to additional spin-related interface resistance $r_s$. This phenomenon is already present on a single interface \cite{vanson_prl1987}. In magnetic multilayers it increases proportionally to the number of interfaces, producing the celebrated giant magnetoresistance effect \cite{baibich:PRL1988,grunberg:PRB1989}.

It was found by Johnson and Silsbee \cite{johnson-silsbee_prl1985}, that accumulation of injected spins can be suppressed by the application of constant magnetic field $H$. They recognized that such suppression is a manifestation of the Hanle effect, originally observed in optical experiments \cite{hanle}.

It was further predicted by Rashba \cite{rashba_ac_injection_APL2002} that when alternating electric current (ac) injects oscillating spin accumulation, the latter lags behind the current and gets suppressed as the current frequency $\omega$ increases. This behavior is described by a complex interface impedance $z_s(\omega)$. Additional effects were predicted when the ac current passes through a spin-valve with two F/N interfaces \cite{fabian_PRL2011}. Alternating current spin injection was a subject of a number of subsequent publications \cite{kaltenborn:PRB2012, wei:NatureComm2014, mangin:APL2017, Hu:IEEE2019, vedyaev:PRB2020}.

The goal of this work is to find what happens when Hanle effect in external magnetic field and the ac drive are combined in one experiment. It will be our result that separate suppressions of spin accumulation produced by each of the two effects do not simply add up but rather create an interesting cooperative pattern.

\section{ac Hanle effect on a single F/N interface}

\begin{figure}[b]
\center
\includegraphics[width = 0.45\textwidth]{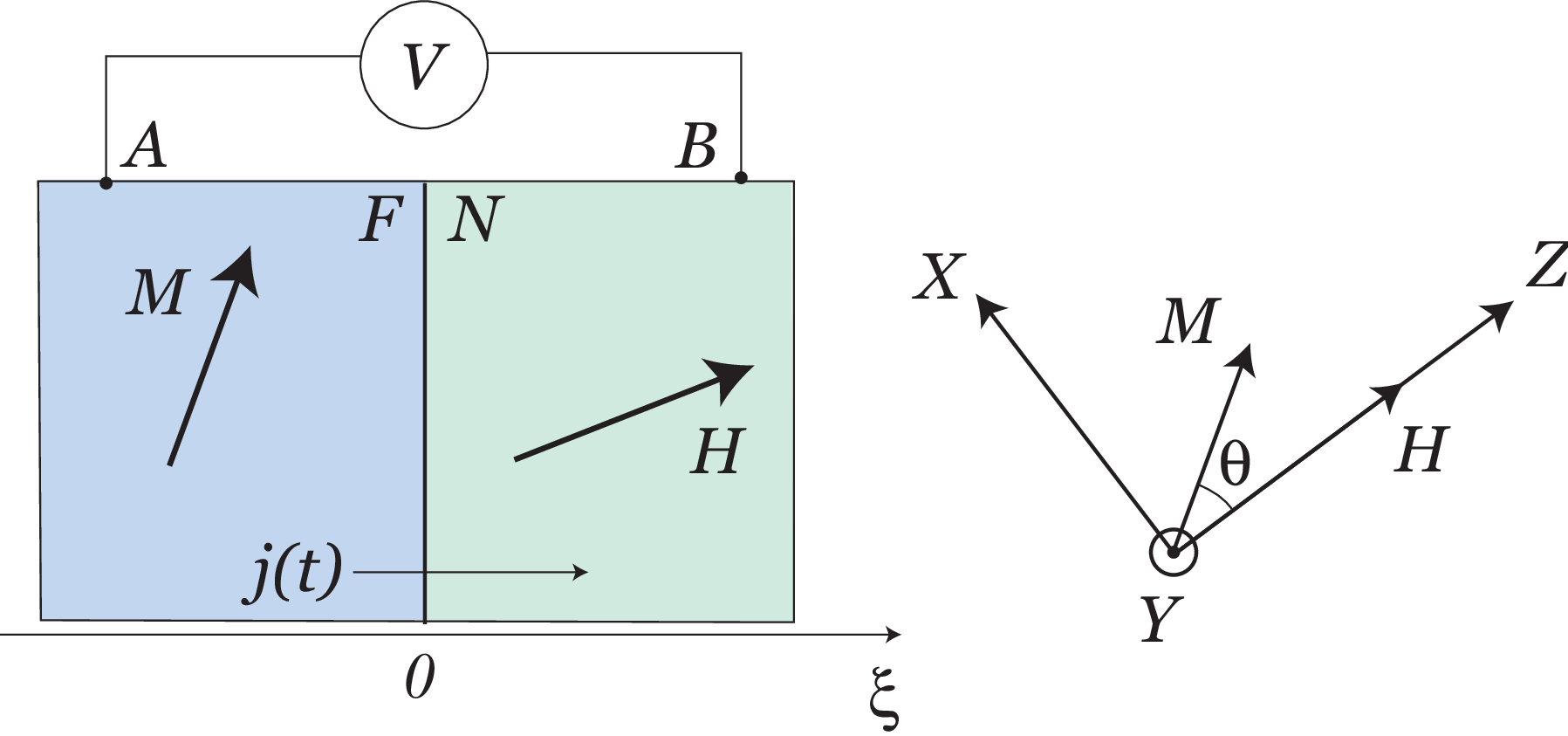}
\caption{Alternating electric current flowing through the F/N interface. Coordinates $(X,Y,Z)$ in spin space are shown on the right. Voltage is measured between points $A$ and $B$.}
 \label{fig:device_sketch}
\end{figure}

We consider a single F/N interface with electric current density $j(t) = j_0 \cos(\omega t)$ flowing perpendicular to it (Fig.~\ref{fig:device_sketch}). Transport is assumed to be diffusive, described by a coupled system known as Valet-Fert equations \cite{campbell:1967,valet-fert_prb1993,rashba_epjb2002, spin-current_book}. All quantities are assumed to depend on just one spatial coordinate $\xi$ perpendicular to the interface. Applied magnetic field $\bf H$ makes an angle~$\theta$ with magnetization $\bf M$ of the ferromagnet. Electron spins interact with $\bf M$ and $\bf H$ through exchange and Zeeman interactions, thus the relative orientation of spin-space axes $(X,Y,Z)$ and spatial axes makes no difference. We choose $Z$ pointing along $\bf H$, and $\bf M$ lying in the $(X,Z)$ plane (Fig. \ref{fig:device_sketch}). It is assumed that magnetization $\bf M$ is not perturbed by the external field, e.g., due to a strong magnetic anisotropy. Furthermore, a standard assumption about spin accumulation in the ferromagnet being always parallel to $\bf M$, even in the presence of external $\bf H$, is also adopted (``strong ferromagnet'' assumption).

Diffusive equations \cite{campbell:1967,valet-fert_prb1993,rashba_epjb2002} are valid at finite temperatures as long as those are much smaller than the Fermi temperature. In metals, this condition is usually satisfied all the way up to room temperature.

In the effectively 1D case considered here, diffusive equations for the spin potential $\vec\mu_s$ and the electrochemical potential $\mu$ decouple \cite{rashba_epjb2002}. In the N-layer we are left with an equation on vector spin potential ${\vec \mu}_{sN}$ written in terms of spin-diffusion length $l_N$ and spin relaxation time $\tau_N$
\begin{equation}
\label{eq:bulk_N_equation}
\dot{\vec{\mu}}_{sN} - \frac{l_N^2}{\tau_N} \partial_{\xi}^2 {\vec\mu}_{sN}  =
    -\frac{\vec\mu_{sN}}{\tau_N} + \gamma {\bf H} \times {\vec\mu}_{sN} \ ,
\end{equation}
where $\gamma = g \mu_B/\hbar$ is the gyromagnetic ratio with material-specific g-factor ($g \approx 2$ in vacuum but can change significantly depending on the band structure), and $\mu_B$ denotes the Bohr magneton. Spin current in N is given by the expression
\begin{equation}
\label{eq:bulk_N_spincurrent}
 {\vec j}_{sN} = - \frac{\sigma_N}{2 e^2} \partial_{\xi} {\vec \mu}_{sN} \ ,
\end{equation}
where $\sigma_N$ is the total electric conductivity of the N-layer and $e < 0$ is the electron charge.

In the F-layer spin potential ${\vec\mu}_{sF} = \mu_{sF} {\vec n}$ points along the magnetization direction ${\vec n} = {\bf M}/M$. The scalar $\mu_{sF}$ satisfies the equation
\begin{equation}
\label{eq:bulk_F_equation}
\dot{\mu}_{sF} -  \frac{l_F^2}{\tau_F} \partial_{\xi}^2 {\mu}_{sF} = -\frac{\mu_{sF}}{\tau_F} \
\end{equation}
with corresponding F-layer parameters, and spin current is given by the expression \cite{rashba_epjb2002}
\begin{equation}
\label{eq:bulk_F_spincurrent}
{j}_{sF} = \frac{P j(t)}{e} - \frac{\sigma_F (1 - P^2)}{2 e^2} \partial_{\xi} {\mu}_{sF} \ ,
\end{equation}
with $P = (\sigma_{\uparrow} - \sigma_{\downarrow})/(\sigma_{\uparrow} + \sigma_{\downarrow})$ being the conductivity polarization in F, where up- and down-spins are characterized by unequal conductivities $\sigma_{\uparrow}$ and $\sigma_{\downarrow}$. Note that the instantaneous value of electric current is entering the expression for $ {j}_{sF}$. This is because momentum relaxation is much faster than spin relaxation, and to a very good approximation charge transport can be viewed as happening in a ``frozen'' spin distribution \cite{rashba_epjb2002}.

Equations (\ref{eq:bulk_N_equation}) and (\ref{eq:bulk_F_equation}) have to be solved with the standard boundary conditions \cite{rashba_epjb2002, spin-current_book}:
\begin{eqnarray}
\label{bc_infty}
&& \mu_{sF} \to 0 \ (\xi \to -\infty), \quad \mu_{sN} \to 0 \ (\xi \to +\infty) \ ,
\\
\label{bc_mus}
&& \vec\mu_{sN}(0) = \mu_{sF}(0) {\vec n}  \ ,
\\
\label{bc_js}
&& ({\vec j}_{sN}(0) \cdot {\vec n}) = j_{sF}(0) \ .
\end{eqnarray}
Condition (\ref{bc_mus}) requires continuity of $\vec \mu_s$ across the boundary, and condition (\ref{bc_js}) expresses the conservation of spin current component along the magnetization. Perpendicular components of spin current are not constrained.

We seek the N-layer solutions of Eq.~(\ref{eq:bulk_N_equation}) in the form ${\vec\mu}_{sN}(\xi,t) = {\vec\mu}_{sN}(\xi) e^{i\omega t} = {\vec A}e^{-\kappa \xi + i\omega t}$ and find the modes satisfying condition (\ref{bc_infty}), namely
\begin{equation}
\label{eq:muN_vs_xi}
\vec\mu_{sN}(\xi) = a_1 \left[\begin{array}{c} 1 \\ i \\ 0 \end{array} \right] e^{-\kappa_1 \xi}
+ a_2 \left[\begin{array}{c} 1 \\ -i \\ 0 \end{array} \right] e^{-\kappa_2 \xi}
+ a_3 \left[\begin{array}{c} 0 \\ 0 \\ 1 \end{array} \right] e^{-\kappa_3 \xi}
\end{equation}
with arbitrary complex amplitudes $a_i$. Wave vectors, expressed through $\omega_h = \gamma H$, are given by
\begin{eqnarray}
\nonumber
\kappa_1 &=& \sqrt{1 + i(\omega + \omega_h)\tau_N}/l_N \ ,
\\
\label{eq:kappa_N}
\kappa_2 &=& \sqrt{1 + i(\omega - \omega_h)\tau_N}/l_N \ ,
\\
\nonumber
\kappa_3 &=& \sqrt{1 + i\omega\tau_N}/l_N \ .
\end{eqnarray}
Similarly, in the F-layer
\begin{eqnarray}
\label{eq:muF_vs_xi}
\mu_{sF}(\xi) &=& a_F e^{\kappa_F \xi} \ ,
\\
\nonumber
\kappa_F &=& \sqrt{1 + i\omega\tau_F}/l_F \ .
\end{eqnarray}

Application of the boundary condition (\ref{bc_mus}) with ${\vec n} = (\sin\theta,0,\cos\theta)$ gives $a_1 = a_2 = (a_F/2) \sin\theta$,  $a_3 = a_F \cos\theta$. Then condition (\ref{bc_js}) leads to the expression for spin accumulation at the interface
\begin{eqnarray}
\label{eq:mu(0)_solution}
&& \mu_{sF}(0) = \frac{2 e P j_0}{S} \ ,
\\
\nonumber
&& S = \sigma_N [(\kappa_1 + \kappa_2)/2] \sin^2\theta +  \sigma_N \kappa_3 \cos^2\theta
\\
\label{eq:S}
&& \quad + \, \sigma_F (1-P^2) \kappa_F \ .
\end{eqnarray}
The latter is known to be directly related to the reading of a voltmeter shown in Fig.~\ref{fig:device_sketch}. It was shown \cite{rashba_epjb2002, fabian:2007, bazaliy:2017} that in the diffusive regime spin-related interface resistance can be understood as a consequence of the presence of
effective distributed electromotive force (e.m.f.) in the F-layer. This e.m.f.\ has volume density $\vec{\cal E} = (P/2e)\vec\nabla\mu_{sF}$, and the extra voltage it produces between the voltmeter attachment points $A$ and $B$ is given by a line integral $V_s = \int_A^B \vec{\cal E} \vec dl$. Since $P = 0$ in N,  the integral should be taken only along the part of the path within the F-layer. In our 1D case this gives $V_s = (P/2e)[\mu_{sF}(0) - \mu_{sF}(A)]$, so if point $A$ is located several spin-diffusion lengths away from the interface, where $\mu_{sF}(A) \to 0$, we get a result that does not depend on the positions of contact points
\begin{equation}
\label{eq:Johnson-Silsbee_voltage}
V_s(t) = \frac{P}{2}\frac{\mu_{sF}(0)}{e} e^{i \omega t}  \ .
\end{equation}
This is, essentially, the Johnson-Silsbee formula \cite{johnson-silsbee_prl1985}, derived in a more general context. As discussed above, $V_s$ depends on the instantaneous value of spin accumulation because momentum relaxation times are negligibly small. Using Eq.~(\ref{eq:mu(0)_solution}) we get $V_s(t) = z_s j(t)$ with spin-related interface impedance (per unit of area)
\begin{equation}
\label{eq:Zs}
z_s(\omega,\omega_h) = \frac{P^2}{S(\omega,\omega_h)} \ .
\end{equation}

\section{Results and Discussion}
Total reading of the voltmeter in Fig.~\ref{fig:device_sketch} is the sum of spin-related voltage and conventional Ohmic voltage $V_{Ohm} = r j(t)$, where resistance $r$ (per unit of area) depends on the placement of contacts $A$ and $B$. The total impedance is given by $z = r + z_s$, with $r$ being independent of spin accumulation.

\begin{figure}[t]
\center
\includegraphics[width = 0.45\textwidth]{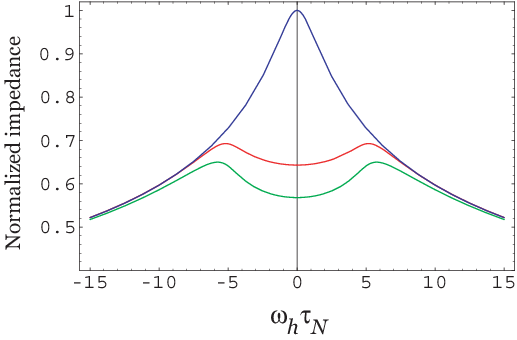}
\caption{Normalized impedances as functions of the rescaled magnetic field $\omega_h \tau_N$ at fixed frequency of the ac current. Blue: pure Hanle effect at zero frequency, $z_s(0,\omega_h)/z_s(0,0)$ ($z_s$ is real at $\omega = 0$). Red: $|z_s(\omega,\omega_h)|/z_s(0,0)$ at finite frequency with $\omega\tau_N = 5$. Green: $Re[z_s(\omega,\omega_h)]/z_s(0,0)$ at $\omega\tau_N = 5$. Other parameters are chosen so that $\sigma_N/l_N = \sigma_F/l_F$, $\tau_F = 0.01 \tau_N$, $\theta = \pi/2$.
}
 \label{fig:spin-resistance}
\end{figure}

If the voltmeter is sensitive to voltage amplitude only, it measures the absolute value of $z$. Since $r$ is not known and depends on the actual experimental setup, we will focus on two limiting cases. First is the case of small Ohmic resistance, $r \ll |z_s|$, with $|z| \approx |z_s|$. Second is the case of large $r \gg |z_s|$, where one can approximate $|z| \approx r + Re[z_s]$ with $r$ playing a role of constant background. Quantities $|z_s(\omega,\omega_s)|$ and $Re[z_s(\omega,\omega_s)]$, norma\-lized on their maximum value $z_s(0,0)$,  are shown in Fig.~\ref{fig:spin-resistance} for a representative choice of materials' parameters that will be discussed below.
Direction of magnetic field is chosen perpendicular to $\bf M$ (i.e., $\theta = \pi/2$) as this ensures the strongest dependence of $z_s$ on the field's magnitude. Results for the cases of large and small $r$ look fairly similar: at finite frequency the top of zero-frequency Hanle curve gets ``cut-off''. Two broad maxima are formed at $\omega_h \approx \pm \omega$, with a shallow minimum between them. Importantly, this shape cannot be described as two sequential suppressions of $|z|$---first by the frequency, and second by the magnetic field. In contrast, at a finite $\omega$ the application of small magnetic field initially increases $|z|$.

\begin{figure}[b]
\center
\includegraphics[width = 0.42\textwidth]{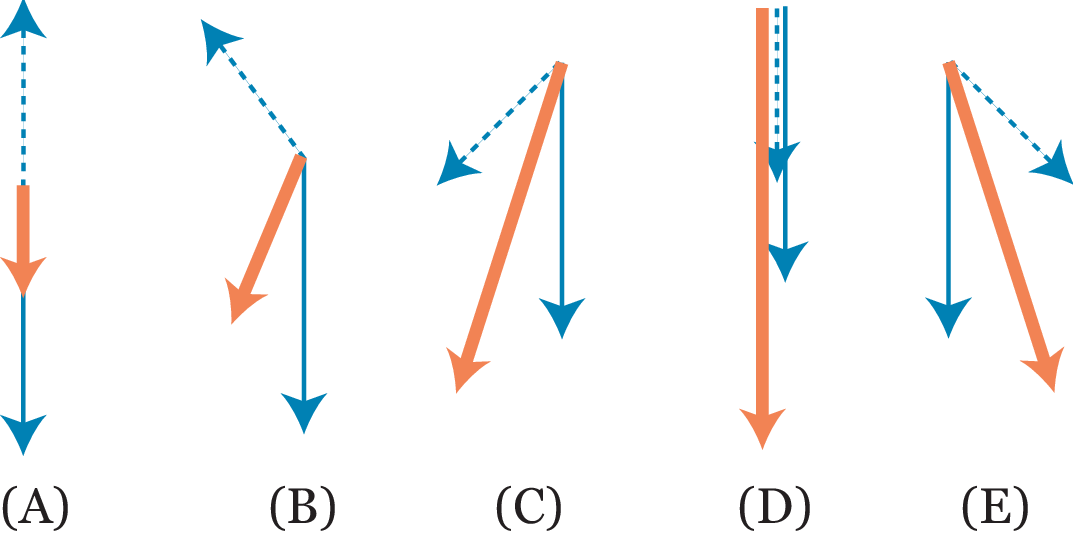}
\caption{Partial spin densities in the N-layer shown for a range of magnetic fields. Dashed blue arrows: partial spin density, injected half a period of ac drive before the observation time, and subjected to decay and rotation by the magnetic field. Full blue arrows: partial spin density injected at the observation time. Red arrows: sums of the two partial spin densities. (A) $\omega_h = 0$. (B) $\omega_h = (1/4) \omega$. (C) $\omega_h = (3/4)\omega_h$. (D) $\omega_h  = \omega$. (E) $\omega_h  = (5/4)\omega$.
}
 \label{fig:physics_picture}
\end{figure}

Such behavior can be rationalized using a physics picture presented in Fig.~\ref{fig:physics_picture}. Spin accumulation at the interface can be understood as an infinite sum of partial spin densities ${\bf s}(t')$, injected at a times $t'$ preceding the observation moment $t$. At the injection moment, each partial density is directed along ${\bf M} || Z$, with a magnitude $s \sim j_0 \cos\omega t'$. After being injected, it decays and rotates in the applied magnetic field. Fig.~\ref{fig:physics_picture} follows two partial densities injected with an interval of half a period of ac drive. First injection happens at $t' = t - T/2$ with $j(t')$ having a maximum positive value. Second injection happens at $t' = t$ with $j(t')$ having a maximum negative value. Figure shows how the sum of these two representative injections increases in magnitude with growing field, reaches a maximum at $\omega_h = \omega$, and then decreases again. This is, of course, only an illustration---the infinite sum of partial spin injections is effectively performed by solving diffusive equations (\ref{eq:bulk_N_equation}) and (\ref{eq:bulk_F_equation}).

Note that in F/N/F structures with two interfaces non-monotonic dependence of dc resistance on magnetic field may be produced by the Hanle effect alone. Such dependence was observed in experiments with metallic nano-structures \cite{jedema:Nature2001} and is predicted to be considerably enhanced in semiconductor structures operating in non-linear regime \cite{yu:PRB2005}. In contrast, non-monotonic behavior is predicted here for a single F/N interface, provided that one switches from dc to ac injection.

The overall scale $z_s(0,0)$ of the spin-related resistance is determined by formula (\ref{eq:Zs}). For normal metal like copper $\sigma_N \sim 10^{8}$~1/Ohm$\times$m, $l_N \sim 100$ nm, $\tau_N \sim 100$~ps \cite{jedema:Nature2001, hamrle:PRB2005}. In ferromagnets, the spin relaxation time and the spin-diffusion length are always much smaller than their counterparts in the normal layer. In permalloy \cite{jedema:Nature2001, hamrle:PRB2005} $\tau_F \sim 0.01 \tau_N$. Permalloy parameters $\sigma_F \sim 10^7$~1/Ohm$\times$m and $l_F \sim 5$ nm correspond to a strong inequality $\sigma_F/l_F \gg \sigma_N/l_N$---an unfavorable regime, where suppression of Hanle peak by the ac current is small. We assume that one can relatively easily decrease $\sigma_F$ and achieve $\sigma_F/l_F \sim \sigma_N/l_N$, while not changing the ratio of relaxation times too much. Conservatively setting $P = 0.1$ for a ferromagnet, one gets $z_s(0,0) \sim 10^{-17}$~Ohm$\times$m$^2$. In a nanowire with cross-section $A \sim (100 \ {\rm nm})^2$ the spin-related resistance will be of the order of $R_s = z_s(0,0)/A \sim 10^{-3}$ Ohm. Figure~\ref{fig:spin-resistance} shows that for an ac current with frequency $\omega \sim 5/\tau_N = 5 \times 10^{10}$ 1/s, or $f = \omega/2\pi \sim 8$ GHz, measuring the red curve will require an accuracy of about $0.01 R_s \sim 10^{-5}$ Ohm. The peak of the curve, assuming $g = 2$, will be located at $H \approx 2.8$~kOe.

\begin{figure}[t]
\center
\includegraphics[width = 0.45\textwidth]{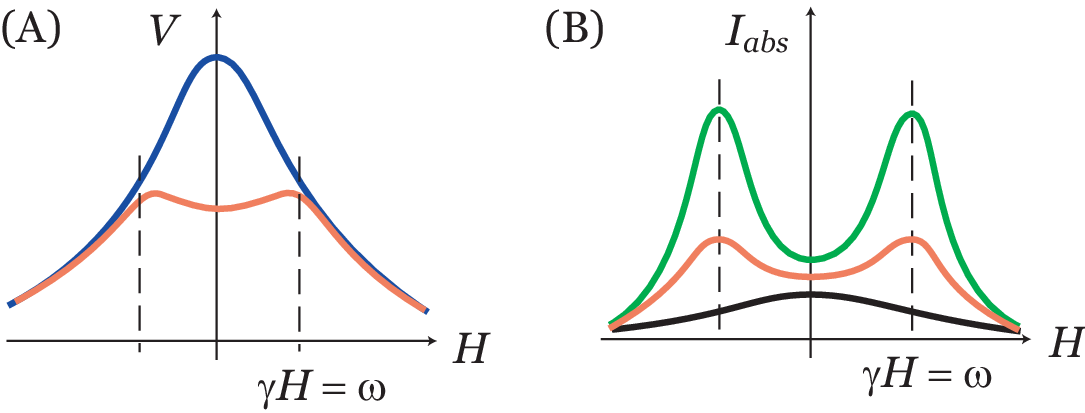}
\caption{(A) Electric response in the spin-injection experiment. Blue line at $\omega = 0$, red line at finite $\omega$. (B) Magnetic resonance absorption signal at finite $\omega$. Green line for small damping, red line for large one, black line for the overdamped regime.}
 \label{fig:acHanle-vs-FMR}
\end{figure}

It is instructive to compare the electric signal from spin accumulation shown in Fig.~\ref{fig:spin-resistance} with the absorption signal measured in a magnetic resonance experiment \cite{kittel-book}. Such comparison is sketched in Fig.~\ref{fig:acHanle-vs-FMR}. One can see that the shape of electric signal is somewhat similar to that of the magnetic resonance signal with large damping. Mathematically, effective damping in the spin injection problem is determined by the real parts of wave vectors (\ref{eq:kappa_N}). Unlike in the magnetic resonance problem, here damping is not a separate parameter, but a function of frequency that remains large in all cases.

From the physics point of view, two situations are simi\-lar in that a) magnetic field $H$ sets the precession frequency $\omega_h$ for spins in the N-layer and b) departure from equilibrium is forced by external periodic drive (radio-frequency field in the case of resonance, and ac electric current in the injection experiment). The difference is in the origin of spin polarization in N: it is created by constant $H$ in the resonance experiment, and by the pre-existing spin polarization in F in the injection experiment. For that reason, injection creates a much larger spin density in the N-layer: indeed, spins in F can be viewed as being polarized by an exchange field of the order of $100 \div 1000 \ T$ \cite{kittel-book}, much larger than $H \sim 1 \ T$ used in the resonance measurements. Competition between large polarization and large effective damping ultimately determines the signal strength that has been evaluated above.

We finally note that in the diffusive regime spin accumulation is not particularly sensitive to interface roughness. Spatial motion of electrons is already chaotic in the bulk, so additional shape imperfections of the interface on the scale of electron mean free path do not change the averages. Imperfections on the scale of spin diffusion length would be important but modern fabrication techniques easily achieve flatness on that scale.

\section{Conclusions}
In the ac Hanle effect one studies the combined action of external magnetic field and alternating current drive on the spin-related resistance of an F/N interface. While separately each factor suppresses such resistance, together they are predicted to lead to a resonance-like effect with the resistance peak observed when the ac driving frequency matches the precession frequency of electron spins in magnetic field.

Such behavior can be viewed as another manifestation of 
complementarity between spin-injection and magnetic resonance that was already emphasized in the original paper of Johnson and Silsbee \cite{johnson-silsbee_prl1985}.

The maxima of spin-related resistance (Fig.~\ref{fig:spin-resistance}) might be even used to measure parameter $\gamma$, and thus determine the material-specific g-factor in the N-layer. Such experiment may be easier to perform than the standard electron paramagnetic resonance measurement, although its accuracy is lower due to the broad shape of the peaks.

\end{document}